# Conductance-dependent Photoresponse in a Dynamic $SrTiO_3$ Memristor for Biorealistic Computing

Christoph Weilenmann[1], Hanglin He[1], Marko Mladenović[1], Till Zellweger[1], Kevin Portner[1], Klemens Bauer[1], Guillaume Bellec[2], Mathieu Luisier[1], Alexandros Emboras[1]

[1]Integrated Systems Laboratory, ETH Zurich, 8092 Zurich, Switzerland.
[2]Machine Learning Research Unit, TU Wien, 1040 Vienna, Austria.

## Abstract

Modern computers perform pre-defined operations using static memory components, whereas biological systems learn through inherently dynamic, time-dependent processes in synapses and neurons. The biological learning process also relies on global signals—neuromodulators—who influence many synapses at once depending on their dynamic, internal state. In this study, using optical radiation as a global neuromodulatory signal, we investigate nanoscale $SrTiO_3$ (STO) memristors that can act as solid-state synapses. We observe that the memristor's photo-conductance exhibits a long-term decay process after photoexcitation (10s of seconds) with an activation energy of 0.33 eV. Based on density functional theory calculations we attribute this long-term photoresponse to the generation and migration of oxygen vacancies at the Pt-$SrTiO_3$ interface. We go on to show that the photo-conductance decay can be accurately controlled through an electrical bias signal and that the magnitude of the memristor's photoresponse depends on its electrical conductance state, following a well-defined square root relation. These properties in combination with our device's low power operation (< 1pJ per optical pulse) and small measurement variability may pave the way for space- and energy-efficient implementations of complex biological learning processes in electro-optical hardware.

## Introduction

Biological neuronal function and synaptic learning mechanisms are governed by decay processes, with a wide range of time constants that depend on the internal dynamics of neurons and synapses [1]–[4]. These dynamics act as a transient memory state that has been hypothesized to help bridge the gap between synaptic and environmental timescales [5], [6]. This so called temporal credit assignment problem is one of the most challenging tasks that biological nervous systems have to solve to be able to learn through reinforcement [5], [7].

Additionally, the synaptic memory state is modulated by local neuronal activity as well as global neuromodulatory signals that influence many synapses at once [8]. These neuromodulators encode environmental cues such as e.g. reward, surprise or novelty, which are critical for the learning process in biology [9]. Integrating these signals with dynamic synapses into neuromorphic hardware could therefore improve the capabilities of bio-realistic, artificial neural networks, especially in dynamic environments [10] where traditional AI approaches lack behind their biological counterpart [11].
One type of such neuromorphic hardware are custom complementary metal-oxide-semiconductor (CMOS) circuits, which have been employed to emulate the complex dynamic behavior of synapses and neurons [12]. For instance, analog circuits based on RC elements, where a capacitor stores charges that are gradually depleted through a resistor, can realize volatile decay processes [13]–[15]. Such implementations offer high controllability and rely on conventional circuit components, making hardware synthesis straightforward [16]. However, this approach also faces limitations, in particular the large footprint of the employed capacitors. Moreover, extending synaptic functions to support more complex plasticity mechanisms, e.g., involving neuromodulators, requires additional circuitry for each synapse, further increasing size and energy consumption. These challenges have prompted the search for alternative hardware paradigms better suited to bio-inspired learning.

Memristor-based architectures show great promise as they can inherently emulate dynamic decay processes [17], [18], [19] without large capacitors or inefficient leakage currents. Memristors are two-terminal nanoscale devices that change their conductance in response to electrical stimuli [20], similar to how synaptic weights vary in response to neuronal spikes. Also memristors sensitive to optical signals have been proposed [21], [22]. Furthermore, studies have demonstrated that, in addition to the device's steady-state conductance, the internal dynamics of memristors can be leveraged to replicate the biophysical processes in synapses and neurons [23]. Indeed, the interplay between electronic, ionic, and thermal processes—each operating on its own energy and time scale—offers the possibility to create a range of different time constants [24]–[28]. By harnessing these dynamic processes, inherent to so-called "dynamic" memristors, bio-realistic synaptic learning rules have been implemented in single devices [24], [29], [30], as well as in synaptic circuits based on memristors [19], [31], [32].
Additionally, the effect of global neuromodulatory signals on conventional (non-dynamic) memristor synapses has been investigated. For example, Sarwat et al. achieved reward-modulated learning by applying combined optical and electrical stimuli to control synaptic changes in phase-change memristive devices [33]. However, due to the use of non-dynamic memristors, the temporal evolution of the internal memory state and its sensitivity to the timing of global modulatory signal delivery could not be explicitly considered. This leaves an opportunity to investigate the combination of a dynamic memristor — that is, a memristor with an evolving memory state — and its interaction with a global neuromodulatory signal. Specifically, we are looking for a space- and energy-efficient hardware implementation of synapses that not only react to local neuronal signals with a change in their memory state, but also allow for higher complexity synaptic functions, which include interactions between a global signal and local, dynamic memory states. Furthermore, the timescale of the state's decay needs to be tuned to the dynamics of the environment or task that the neural network implements [31]. This requires close control over the temporal evolution of the memory state for each individual synapse.

In this paper, we present a dynamic memristor that realizes these functions. Namely, it (1) responds to local signals by a short-term change in its memory and then decays back (dynamic memory state), (2) is sensitive to global signals in a way that depends on that dynamically evolving local state (state-dependent global modulation), and (3) allows for exact control over the state's temporal evolution (controllable decay time). Specifically, we fabricated two-terminal Pt/Cr – $SrTiO_3$ – Ti/Pt memristors, whose conductance models the synapse's dynamic memory state that interacts both with electrical (local) and optical (global) signals. In this scheme, light acts as a neuromodulator, which offers several advantages: (i) It is not limited by the RC time constant of electrical stimuli (low latency) [34], (ii) it can address many devices simultaneously, without complex wiring and associated peripheral circuits (small spatial footprint), and (iii) it consumes < 1pJ per optical pulse (high energy efficiency). Altogether, these features allow for the implementation of global neuromodulator stimulations on a small spatial footprint and at low energy consumption. A combined density functional theory (DFT) and experimental investigation of the physical processes underlying the photoresponse of our memristor was also conducted. The results suggest photogenerated oxygen vacancies and their subsequent migration as a likely mechanistic explanation for the experimentally observed device behavior.

We go on to show that the photoresponse to the global neuromodulator depends on the memristor's conductance value (i.e., its memory state) and that the temporal dynamics of our device's conductance state can be controlled through the introduction of an electrical bias signal. Finally, we link these findings to biological learning processes, demonstrating how our device properties can be used in so called three-factor learning rules [8]. These rules could expand the capabilities of bio-realistic artificial neural networks to new tasks such as sequential learning [35], reinforcement learning [36], [37], and classification in dynamic environments [38], potentially providing a more energy-efficient alternative to traditional neural networks [39]. The space- and energy-efficient hardware implementation of such higher-complexity synaptic learning rules highlights the potential of our electro-optical devices as building blocks for future neuromorphic circuits.

## Results and Discussion

### Experimental Setup and Basic Device Working Principle

The experimental setup and device schematic are illustrated in Figure 1a. The device can be measured electrically and illuminated with UV light (light-emitting diode with $\lambda$ = 365 nm) from top through an optical fiber. The specific wavelength was chosen, because the photon energy $h\nu = 3.4$ eV lies above the bandgap of $SrTiO_3$ (3.25 eV [40]) and therefore stimulates interband transitions. The memristor consists of two electrodes: Pt(50 nm) / Cr(2 nm) and Ti(35 nm) / Pt(10 nm) that were each deposited in two separate steps on a single-crystal $SrTiO_3$ (STO) substrate (10 x 10 x 0.5 mm from SurfaceNet) through e-beam evaporation. The surface of the STO substrate is in (100) orientation with a mixed $TiO_2$ and SrO termination. The electrodes were patterned using e-beam lithography and a subsequent lift-off process, resulting in a gap between the electrodes of approximately 40nm. After patterning, the device was subjected to an annealing treatment (300°C in Ar atmosphere for 20 min.) to improve electrical contact between the electrodes and the STO substrate. In this step a thermal oxide is likely formed below the Ti electrode [41]. The whole device is encapsulated by a SiN layer deposited via plasma enhanced chemical vapor deposition (PECVD) that acts as an oxygen and moisture barrier. A FIB/SEM cross-section of the fabricated device is presented in Figure 1b.

The current-voltage characteristics of the device at different optical illumination intensities are shown in Figure 1c. Under dark conditions (black line), when the voltage is swept from -2V to 2V, the current exhibits the typical "eightwise" hysteresis [42] of an STO memristor. This hysteresis is mainly attributed to the modulation of the Schottky barrier due to the creation and migration of oxygen vacancies ($V_O^{\bullet\bullet}$) near the Pt/Cr/STO interface [43], [44]. When light is turned on, the current-voltage characteristics shift towards higher currents. While this light-induced increase of the electrical conductance usually disappears immediately after turning off the optical signal [45], in our devices, it only slowly decreases back to its initial value, as can be seen in Fig. 1d. In this experiment the current is measured at a constant read voltage of $V_{read} = 0.6$V, while UV pulses of different durations are applied. Under illumination, the current steadily increases, which leads to a higher photocurrent for longer pulses. The initial relaxation of the photocurrent (timescale of ms) is more pronounced for longer pulses, while the long-term decay (10s of seconds) is similar for all pulse widths.

Figure 1e reveals how the decay of the current under optical excitation depends on the device conductance state. As a reference, we first consider a purely electrical response (black and grey lines), where the mean (solid lines) and standard deviation (shaded area) over three measurements of the same device are reported for two distinct initial conductance states ($G_o^{(i)}$ and $G_o^{(ii)}$). In both cases, an electrical pulse with amplitude $V_{pulse} = 1.9V$ is applied during 100 ms, which leads to a conductance increase, followed by a decay during which the current is read at $V_{read} = 0.6$V. The experiment is then repeated using an optical instead of an electrical pulse (optical pulse with $P_{opt} = 65mWcm^{-2}$ and a duration of 100 ms; purple curves). For the $G_o^{(i)}$ state, the electrical and optical responses exhibit a very similar time decay. However, for the $G_o^{(ii)}$state, the conductance is significantly higher in the optical than in the electrical case immediately after the pulse.

The origin of electrically-induced conductance decay in STO-based memristors is a topic of active research, possibly involving different physical mechanisms: It has been attributed to (i) the back-migration of oxygen vacancies toward the Pt-STO interface and subsequent filling of vacancies by oxygen incorporation [46], [47], (ii) the slow release of trapped charges [48], or (iii) a combination of both effects [49]. Purely electrical measurements are not sufficient to unambiguously determine the phenomena governing the observed conductance decay as measured data can be fitted with both oxygen vacancy migration and trapping/detrapping models [50], [51]. Nevertheless, experimental evidence has been accumulating that underlines the sensitivity of the conductance decay on the oxygen partial pressure during measurement [47], [52], the presence of oxygen blocking layers [53], [54], and the oxygen permeability of the top electrode [30]. Furthermore, the generation of oxygen vacancies upon electrical stimuli has been experimentally verified through electron energy-loss spectroscopy [43]. Such experiments support an ionic origin of the conductance decay rather than a pure trapping / detrapping mechanism, at least for sufficiently high pulse voltages.

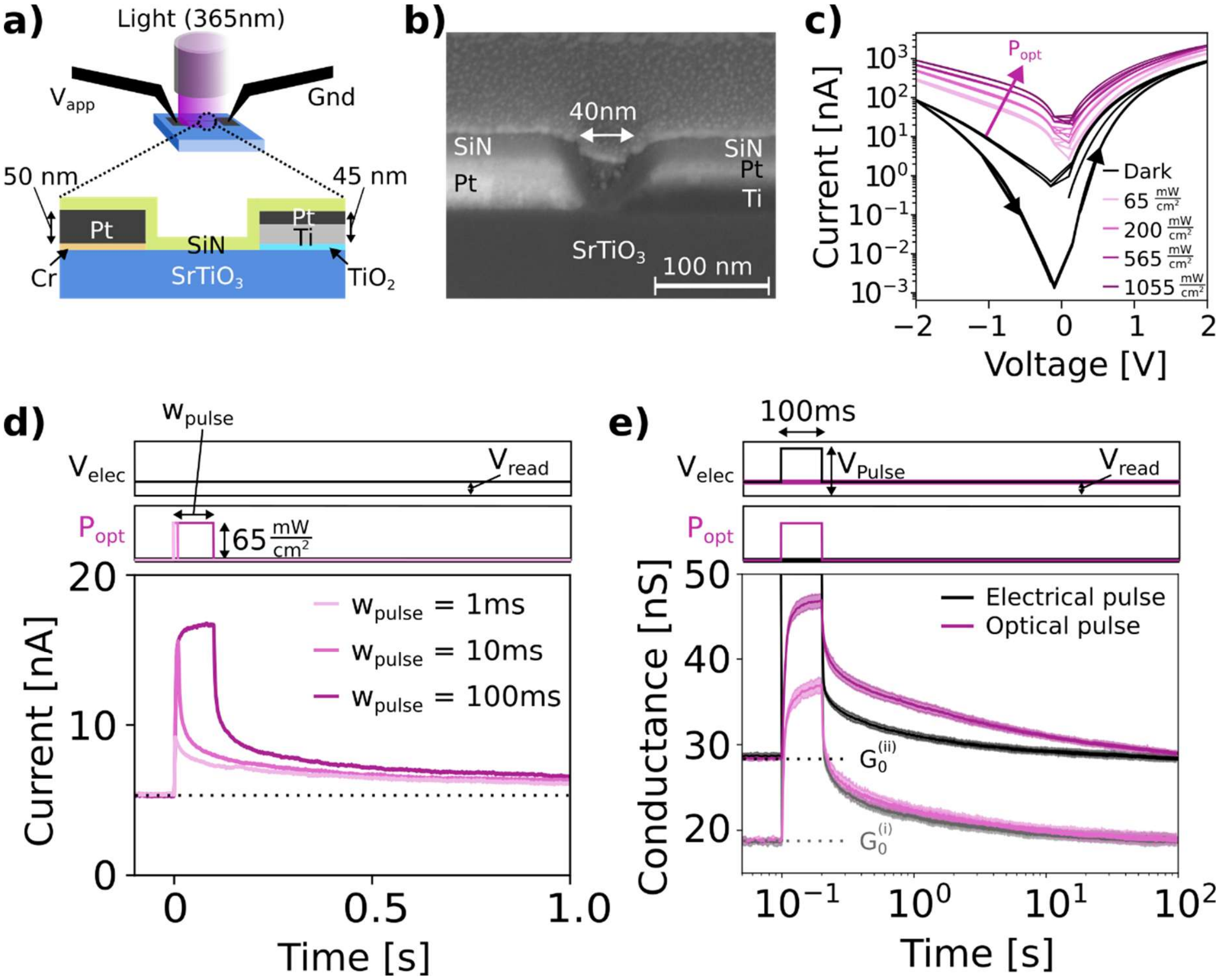

Figure 1: Experimental setup, device design and basic characterization of our Pt/Cr – $SrTiO_3$ – Ti/Pt memristor. (a) Schematic illustration of the experimental setup and of the device stack. Free-space illumination at a wavelength of $\lambda = 365$ nm is focused on the active region of our two-terminal STO-based memristive device. A voltage can be applied to the Pt/Cr electrode, whereas the Ti/Pt one is grounded. (b) Cross-sectional SEM image of the memristor gap. The topmost material is a protective Pt layer. It is used to fabricate the cross-sectional slice, but is not part of the device itself. (c) IV characteristics of the device, measured in the dark (black lines) and under varying light intensities ($P_{opt}$, indicated by shaded purple regions). (d) Current as a function of time measured at a constant read voltage $V_{read} = 0.6$V. At t=0 optical pulses with a power of 65 $mWcm^{-2}$ and different widths are applied. (e) Conductance as a function of time under the application of both electrical (black lines) and optical (purple lines) pulses at two different initial conductance states ($G_0^{(i)}$ and $G_0^{(ii)}$). The solid lines denote the mean of 3 measurements and the shaded area around the lines corresponds to the standard deviation.

The similarity between the optical and electrical responses of $G_o^{(i)}$ in Fig. 1e suggests that their decay behavior depends on the same mechanism. Based on the previous discussion, this mechanism could be: (1) the generation and subsequent back-filling of oxygen vacancies with oxygen or (2) charge trapping and subsequent slow release of trapped carriers. Indeed, long relaxation times have been observed for the photoresponse of STO. They have been attributed to interface charge trapping [55], trapping at bulk defects [56], [57], or light-induced generation of oxygen vacancies [58], [59]. While UV-light-induced generation of oxygen vacancies in undoped STO has, to our knowledge, only been reported at higher optical intensities than in our work (2.4 $W/cm^2$ compared to our 96 $mW/cm^2$) [59], experiments with p-type STO at elevated temperatures (> 250°C) have measured a change in oxygen stoichiometry upon illumination with low-intensity UV light (240 - 370 $mW/cm^2$) [60]–[62]. Furthermore, it has been observed, that the formation energy of oxygen vacancies in STO decreases significantly in the proximity of high work-function metals (e.g. Pt [44], [63]), making the light assisted generation of $V_O^{\cdot\cdot}$ at the Pt-STO interface of our devices feasible. On the other hand, charge trapping has also been shown to lead to persistent photoconductivity in STO, but experimental [57] and theoretical [64] works have traced back the origin of this phenomenon to substitutional hydrogen defects, which form only under high temperature annealing (420 – 1200 °C) in the presence of hydrogen-containing compounds [57], [65]. By contrast, analyzing the UV photoresponse of various untreated, commercially available single-crystal STO substrates revealed no persistent photoconductivity [66], effectively ruling out trapping at bulk defects (e.g. due to impurities) as an explanation for the prolonged decay of the photoresponse in our devices. Charge trapping effects in the SiN layer can also be ruled out (see Supplementary Section S1, Fig S1). Nevertheless, it is conceivable that interface defects (intrinsic or introduced during fabrication) could form trap states affecting the photoresponse.

**A Possible Physical Mechanism Underlying the Photoresponse**

To investigate the physical origin of the slowly decaying photoresponse the Pt-STO interface is analyzed during and after illumination. The aim is to evaluate the evidence for the two different physical processes ($V_O^{\cdot\cdot}$ generation + migration and charge trapping/detrapping) using a combination of experimental and simulation results. To that end Fig. 2 is considered, where the physical processes relevant for $V_O^{\cdot\cdot}$ generation are numbered (1) to (5). Analogously, the processes relevant for charge trapping are discussed in Supplementary Section S2 and visualized in Supplementary Fig. S2. Figure 2a shows the proposed band diagram of the interface before (black) and during illumination (pink). (1) During illumination with a wavelength corresponding to an energy above the band gap electrons are excited from the valence band into the conduction band. (2) The photogenerated holes migrate towards the Pt contact, where they accumulate due to the limited surface recombination velocity at the non-ideal interface [67]. An atomically thin insulating layer is drawn at the interface between Pt and STO to model this non-ideal contact. The accumulated holes cause an increased band bending at the interface, which leads to a thinner Schottky barrier. This decrease in Schottky barrier thickness and the additional conduction band electrons lead to a larger electron current $J_n$ during illumination, illustrated as arrows on top of the band diagrams. This increased current is responsible for the conductance increase during an optical pulse. To verify if holes in the valence band promote the generation of oxygen vacancies we carried out Constrained Density Functional Theory (CDFT) simulations (see bottom of Fig. 2a), which allows for defining the electron population on specific atoms (see Supplementary Section S3 for details). (3) By localizing two holes at an interfacial Oxygen 2p orbital (which form the valence band maximum in STO) the Ti-O bond is broken. (4) This bond breaking allows for the Oxygen to move to an interstitial position, leaving behind a $V_O^{\cdot\cdot}$. At the interface, the enthalpy of the full reaction (steps (3) and (4)) is negative (i.e. exothermic) when photogenerated holes are localized on the oxygen atoms (see Table 1). This suggests that $V_O^{\cdot\cdot}$ generation is a spontaneous process at the Pt-STO interface under UV light illumination. When the light is turned off (Figure 2b) the band structure near the interface has been modified by the photogenerated oxygen vacancies even after all holes have recombined. This non-steady state allows for a larger electronic current (higher conductance) than the steady state (dashed lines in Fig 2b), due to additional conduction band electrons and a thinner electron tunneling barrier [68]. In this picture the slow conductance decay after an optical pulse is due to (5) the back-migration of oxygen vacancies to the interface, driven by the built-in electric field in the depletion region of the Schottky junction. We have calculated the activation energy for this vacancy migration, as illustrated in the bottom part of Fig 2b, to be 0.34 eV (see Table 1).

| Process | Energy [eV] |
|---|---|
| $V_O^{\cdot\cdot}$ generation (dark) | |
| Pt-SrO interface | 0.83 |
| Pt-TiO2 interface | 3.84 |
| $V_O^{\cdot\cdot}$ generation (light) | |
| Pt-SrO interface | -2.21 |
| Pt-TiO2 interface | -1.01 |
| $V_O^{\cdot\cdot}$ migration (bulk) | 0.34 |

Table 1: DFT results for the energy requirements of physical processes involved in the generation and migration of oxygen vacancies. The values for $V_O^{\cdot\cdot}$ generation in the dark and for migration are activation energies of these processes. The values for $V_O^{\cdot\cdot}$ generation under illumination represent the enthalpy of the reaction (i.e., the difference between the initial and final state).

To verify if this mechanism is consistent with our experiments, we performed optical pulse measurements at different temperatures (Figure 2c) and observed the timescale of the photoresponse decay. Here, the current is plotted as a function of time after an optical pulse (100ms / 65mW/cm$^2$) was applied. To better compare the decay of the different measurements the data was shifted, such that the value for the current before the pulse is equal to zero (i.e., $10^{-\infty}$ in the log plot) for all temperatures. The dotted lines denote fits with a double exponential model that is parametrized by two time constants: $\tau_1^{(opt)}$ for the initial fast decay and $\tau_2^{(opt)}$ for the long term decay. The values of these time constants are displayed in the Arrhenius plot in Fig. 2d. For the long term decay this measurement reveals an activation energy of 0.33 eV, which is in excellent agreement with the calculated value for $V_O^{\cdot\cdot}$ migration. Contrary, the calculations and discussion on the process of interface charge trapping in Supplementary Section S2 suggest timescales much shorter than observed in our experiments. We therefore attribute the long term decay to the back-migration of photogenerated $V_O^{\cdot\cdot}$ and the short term decay to a combination of $V_O^{\cdot\cdot}$ migration and charge trapping. We note however, that we have not modelled all possible interface states, just interfacial oxygen vacancies and can therefore not completely rule out the effect of charge trapping on the long-term decay.

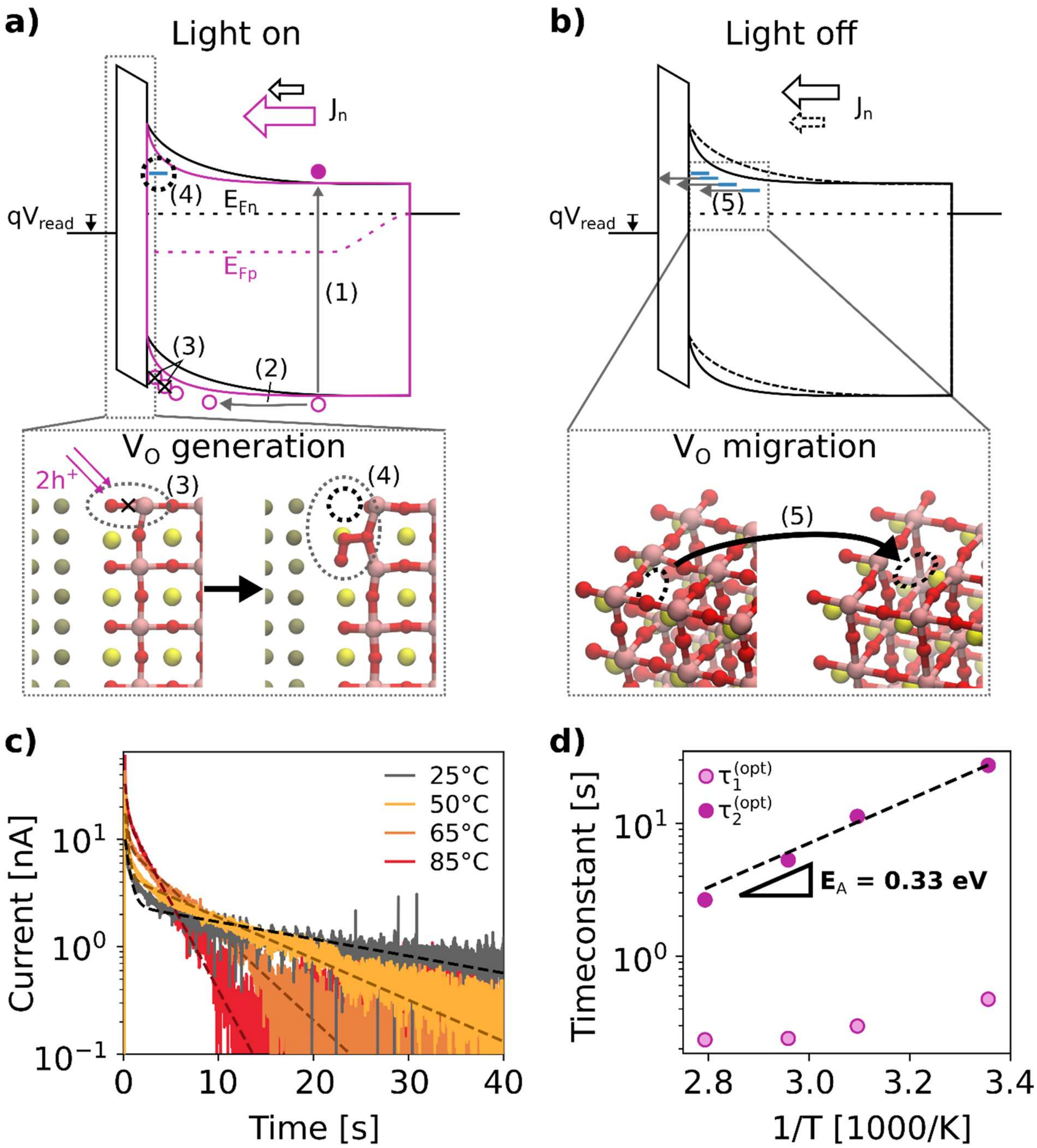

Figure 2: Investigation of the physical mechanism behind the photoresponse. Processes that are relevant for $V_O^{\cdot\cdot}$ generation are marked with numbers (1) to (5). (a) On top the band diagram of the Pt-STO Schottky interface before (dark) and during (pink) illumination is displayed. The electron current $J_n$ is shown above the band diagram. Below the DFT simulation for optically induced $V_O^{\cdot\cdot}$ generation at the Pt-STO interface (SrO terminated) is shown. The atoms displayed are Pt (grey), Sr (yellow), Ti (pink), O (red) (b) On top the band diagram immediately after illumination (continuous line) and in steady state (dashed line) is shown. Generated oxygen vacancies are shown as blue in-gap states. Below a DFT simulation of the vacancy migration process is displayed. (c) Current as a function of time read out at 0.6V after an optical pulse (100ms / 65mW/cm$^2$) for different temperatures. For each temperature a double exponential fit is shown with the dashed lines. (d) Arrhenius plot of the time constants extracted from the fit in c). An activation energy of $E_A$ = 0.33eV was obtained for the long-term time constant.

**Conductance-dependent Photoresponse**

To further investigate the conductance-dependent photoresponse of our memristor, we illuminate it with UV pulses (1 ms, $P_{pulse} = 65\ \mathrm{mW/cm^2}$) before and after an electrical SET at t=0 (Fig. 3a), while measuring the conductance at a read voltage of $V_{read} = 0.6V$. During the SET procedure a train of 100 electrical pulses is applied ($V_{SET} = 4\mathrm{V}, 1\ \mathrm{ms}$). The electrical and optical protocol are shown at the top of the sub-plot, the measured conductance being plotted at the bottom. Two measurements are displayed that differ in the timing of the UV pulses after SET (light and dark blue curves). The conductance measurement begins with a slow transient phase that lasts until t=-400s, when the first optical pulse is applied in both measurements. A short-term conductance increase is observed whenever the device is illuminated, followed by a slow decay back to the initial value. The near-perfect overlap of the two measured conductance traces over the first 500s shows the low measurement variability of our device. Furthermore, the optical energy consumption per pulse is estimated to be 0.37 pJ (see Supplementary Section S4), assuming a crossbar array architecture with a realistic device pitch area of 0.5625 $\mu m^2$, as demonstrated in [69]. The electrical power consumed during the read operation at $V_{read}$=0.6 V ranges between 2 and 22 nW, depending on the conductance state. Additional electrical measurements of the same device confirm a low cycle-to-cycle variability [70].

After the electrical SET, the conductance increases roughly by a factor of 6× (see Fig. 3b), with a subsequent decay in the timescale of several 100's of seconds. This long-term change in conductance enables us to compare the photoresponse at various conductance states corresponding to different measurement segments, labeled (1), (2), and (3). Figure 3c highlights these three segments for one of the two measurements (measurement 2). Each of the segments starts at a distinct initial (transient) conductance state ($G_0^{(1)}$, $G_0^{(2)}$, and $G_0^{(3)}$), which leads to different photoresponses ($\Delta G^{(1)}$, $\Delta G^{(2)}$, and $\Delta G^{(3)}$) under optical excitation. The following trend is observed: the larger the conductance state, the larger the photoresponse. This behavior is consistent with the proposed mechanism of optically induced oxygen vacancy generation: A larger conductance state corresponds to more band bending, which in turn favors the accumulation of holes that promote the formation of oxygen vacancies. As the conductance of the device naturally decreases over time, the strength of the photoresponse also decreases as the UV pulse is delayed with respect to the electrical SET. In other words, the closer an optical pulse occurs to the SET event, the larger the resulting photoresponse. This timing dependence enables the implementation of transient memories of past activity that persist until a reward signal (optical pulse) arrives. Such transient memories are a critical element of three-factor learning rules, as they allow synapses to link local events with delayed global feedback [5].

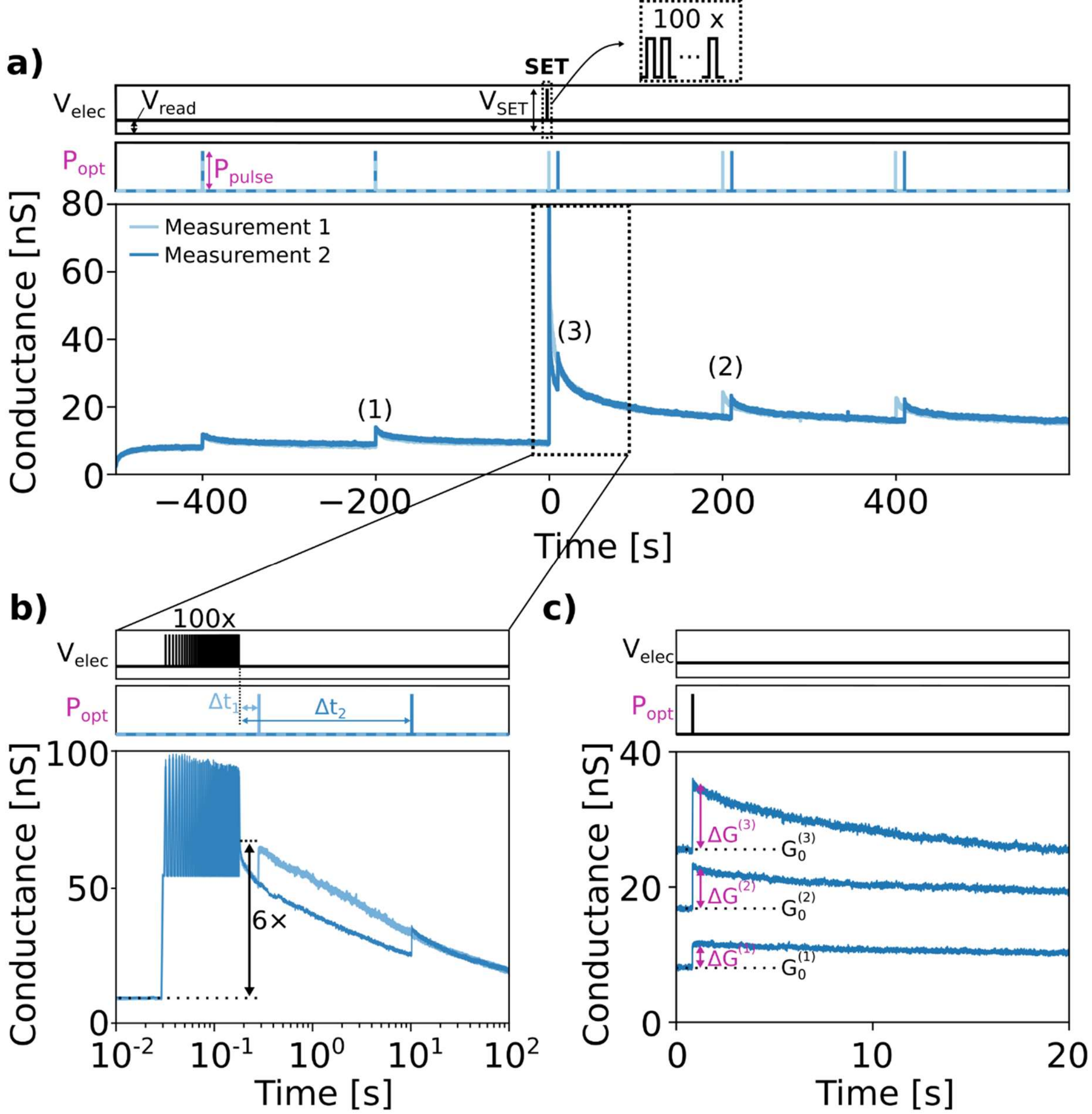

Figure 3: Characterization of the electrically and optically induced conductance change of our STO dynamic memristor. For each measurement the electrical protocol ($V_{elec}$) and the applied optical power ($P_{opt}$) are shown in the top two panels of each subfigure. (a) Two separate measurements (light and dark blue) of the conductance vs. time characteristics differing in the timing of the applied UV pulses after electrical SET (see the protocol on top). (b) Zoom-in of the measurement region in the dotted rectangle plotted on a logarithmic x-axis. The timing difference of the optical pulses with respect to the end of the electrical SET are $\Delta t_1 = 0.1$s and $\Delta t_2 = 10$s for measurement 1 and 2 respectively. (c) Plot of the measurement parts labelled (1), (2) and (3) in (a). The initial conductance state $G_0^{(1,2,3)}$ before the arrival of an optical pulse and the photoresponse $\Delta G^{(1,2,3)}$ are indicated.

Besides the state-dependent (or timing-dependent) magnitude of the photoresponse, its temporal evolution and its controllability are of great interest as they can be leveraged in a wide range of synaptic learning rules [8], [30], [71], [72] and machine learning algorithms [70], [73], [74]. To this end, we execute the measurement protocol shown in Fig. 4a (top panel) and report the resulting conductance in its bottom panel. Here, contrary to the measurement in Fig. 3a, we do not set a constant read voltage, but instead read pulses (0.5ms, $V_{read} = 0.6$V) with a variable bias voltage in-between (1.5ms, $V_{bias}$). This protocol allows to read out the conductance at a fixed voltage ($V_{read}$), while simultaneously changing the time-averaged voltage applied to the device ($\overline{V_{app}} = (0.5 \cdot V_{read} + 1.5 \cdot V_{bias})/2$ ), with $V_{bias}$ as control parameter. We can thus observe the decay behavior for different $\overline{V_{app}}$ in distinct measurements, while still reading out the conductance in a consistent way (i.e., at the same voltage) for all measurements. The effect of the bias voltage on the conductance is threefold: (1) During the initial transient phase ($t < -400$ s) the conductance increases for $\overline{V_{app}} > 0$ (i.e., at a bias of 0 V and 0.6 V) and decreases for $\overline{V_{app}} < 0$ (at a bias of $-0.6$ V). The conductance change is larger for larger values of $\overline{V_{app}}$. (2) The decay of the conductance state $G_0(t)$ after the electrical SET ($t > 0$) is influenced by the bias voltage (a higher $V_{bias}$ leads to a slower decay). Finally, (3) the decay after the optical pulses is similarly affected.

We first focus on the decay after the SET, i.e., the electrically-induced conductance change. It can be fitted by combining one exponential and one stretched exponential function (dashed black line in Fig. 4a). This SET decay model is also shown in the inset as a log-log plot. Its behavior is captured by the following equation

$$G_0(t) = G_{steady} + \frac{\Delta G^{(SET)}}{\gamma^{(SET)} + 1} \cdot \left( \gamma^{(SET)} * \exp\left( -\frac{t - t_0}{\tau_1^{(SET)}} \right) + \exp\left( -\left( \frac{t - t_0}{\tau_2^{(SET)}} \right)^{\beta} \right) \right). \tag{1}$$

Here, $G_{steady}$ is the initial conductance state before the SET pulses, $\Delta G^{(SET)}$ denotes the difference in conductance between $G_{steady}$ and right after the SET pulses (i.e., the conductance increase), $\gamma^{(SET)}$ is a fitting parameter that weights the relative strength of the first and second exponential, $\beta$ is the stretching exponent, $\tau_1^{(SET)}$ and $\tau_2^{(SET)}$ are time constants, and $t_0$ corresponds to the time immediately after the electrical SET pulses. The values for the parameters are given in Supplementary Section S5. The SET decay involves at least two time constants: a small one dominating at the beginning of the decay process ($\tau_1^{(SET)}$) and a larger one for the remainder ($\langle \tau_2^{(SET)} \rangle$). Therein, $\langle \tau_2^{(SET)} \rangle$ refers to the mean time-constant of the stretched exponential function. It is defined as

$$\langle \tau_2^{(SET)} \rangle = \frac{\tau_2^{(SET)}}{\beta} \cdot \Gamma\left( \frac{1}{\beta} \right) \tag{2}$$

where $\Gamma(\cdot)$ represents the mathematical gamma function. Similar fitting models have been reported in the literature for the conductance decay in STO and other oxide memristors, including stretched exponentials [75], [76] and a combination of exponentials [46], [77], [78]. Also, power laws have been proposed [48], [79] as well as combinations of stretched exponentials with different time constants [80] or power laws and exponentials [51].

As a second step we characterize the decay after the optically induced conductance change at different bias voltages. It can be observed in the zoom-in of the optical pulses at t=-200s in the dotted rectangle (Fig. 4b). Before these pulses, the conductance reaches a steady state for all three measurements ( $G_0(t) = G_{steady}^{(V_{bias})}$ ), allowing for the observation of the relaxation dynamics of the photoresponse ($G_{opt}(t)$) at a stable conductance state. The purple dashed lines corresponds to a fit with a double exponential model and two time constants $\tau_1^{(opt)}$ and $\tau_2^{(opt)}$

$$G_{opt}(t) = G_0(t) + \frac{\Delta G^{(opt)}}{\gamma^{(opt)} + 1} \cdot \left( \gamma^{(opt)} * \exp\left( -\frac{t - t_0}{\tau_1^{(opt)}} \right) + \exp\left( -\frac{t - t_0}{\tau_2^{(opt)}} \right) \right) \tag{3}$$

The values of all parameters can be found in the supplementary material. Therein, also Fig. 4a is replotted with the conductance state $G_0(t)$ subtracted (Supplementary Fig. S3 in Section S6) to visualize the optical decay separate from the electrical decay. The electrical and optical time constants are plotted as a function of the bias voltage in Fig. 4c, indicating that the time constants increase at larger $V_{bias}$. This observation is consistent with the proposed physical mechanism in Fig. 2, as the larger positive (negative) bias hinders (facilitates) the back migration of oxygen vacancies. We have thus shown that it is possible to control the decay timescale of both electrically and optically induced conductance changes through the application of an external bias voltage.

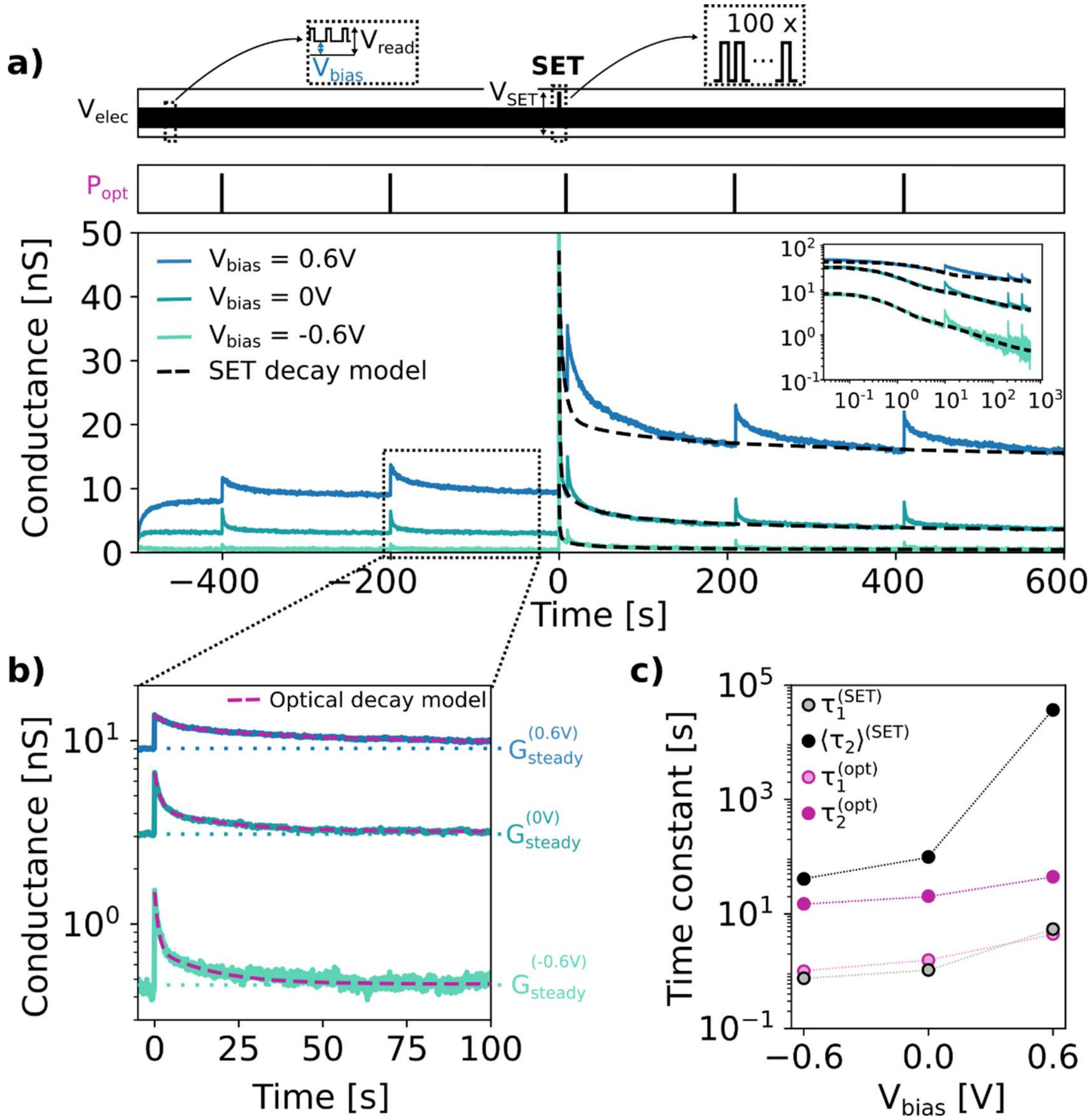

Figure 4: Controllability of the conductance decay rate after the application of electrical and optical pulses. (a) (Top) Applied voltage protocol using 0.6V (0.5ms) read pulses and a variable bias voltage in-between (1.5ms). A train of 100 SET pulses (4V, 1ms) is applied at t=0. The applied optical pulses (1ms, 65mW/cm$^2$) are shown in the panel below. (Bottom) Conductance values at $V_{read}$ =0.6V for three different bias voltages ($V_{bias}$). The relaxation after the SET is fitted for each $V_{bias}$ with the decay model in Eq. (1) (dashed black lines). The inset shows the same data for t > 0 on a log-log plot. (b) Zoom-in of the area marked by the dotted rectangle in (a). The three different initial conductance states ( $G_{steady}^{(V_{bias})}$ ) are labelled and the relaxation of the photoresponse is fitted with the optical decay model in Eq. (3) (purple dashed lines). (c) Extracted decay time constants from the SET and optical decay models as a function of $V_{bias}$. The dashed lines serve as guides to the eye.

Finally, to determine the dependence of the photoresponse $\Delta G^{(opt)}$ with respect to the device conductance $G_0$, we report both quantities in Fig. 5a for a large set of measurements. Two different SET protocols were used: one employing single electrical pulses to SET the device and one where a train of 100 pulses is applied (see Supplementary Section S7, Fig. S4 for details). In Fig. 5a, the markers' color denotes the time interval between the electrical SET and the subsequent UV pulse, while their shape distinguishes different bias regimes. As indicated by the dashed line, the photoresponse $\Delta G^{(opt)}$ increases with $G_0$, with the exception of a few outliers (dotted ellipses). Apart from these outliers the relation between $\Delta G^{(opt)}$ and $G_0$ follows a power law:

$$\Delta G^{(opt)} = G_0^k$$

*( 4 )*

with an exponent $k$ = 0.50 (i.e., a square root relation). Neither the applied protocol (marker size), the UV pulse timing with respect to the electrical SET (color), nor the applied bias voltage (marker symbol) affect the observed power law. We therefore conclude that the conductance state of the device ($G_0$) is the main predictor for the magnitude of the photoresponse ($\Delta G^{(opt)}$). Figure 5b shows the photoresponse as a function of the optical power for a predefined conductance state. We observe larger $\Delta G$ at higher optical power, with a power law exponent $k$ = 0.52.

## Bio-inspired Learning Rules

Our memristor enables the interaction of light with the device's dynamically decaying conductance state. This gives rise to a conductance-state-dependent photoresponse that is sensitive to the relative timing of the optical signal. This feature is, to the best of our knowledge, reported here for the first time and could potentially be used by advanced neuromorphic computing circuits.

In the following, we argue that the combination of features of our device: (1) dynamic memory state, (2) state-dependent global modulation, and (3) controllable decay enables the implementation of bio-inspired three-factor learning rules [8]. In short, the light beam could be used to communicate a reward signal globally across a cross bar array while the dynamic conductance state implements the synaptic dynamics physically (see Fig. 5c). The following paragraphs describe a possible match between a simple form of reward based 3-factor rules and our device dynamics. We also discuss the limitations of this analogy.

**Background on Three-Factor Learning Rules.** 3-factor learning is a theoretical framework that is intended to model synaptic plasticity. For a review of the biophysical data supporting 3-factor rules we refer to [1]. One example of such a rule is reward-modulated spike timing dependent plasticity (R-STDP), which governs how the synaptic weight ($\mathrm{w_t}$) is updated according to its local pre- and postsynaptic inputs ($pre$ and $post$) and to a global neuromodulatory reward signal ($\delta_t$):

$$w_t = w_{t-1} + \Delta w_t(pre, post, \delta_t).$$

*( 5 )*

In real world environments the global reward signal arrives much later than the local neuronal activity ($pre$, $post$) that led to the rewarded behavior. To bridge this temporal gap, it is crucial that synapses hold a local transient memory of previous activity with timescales adapted to the specific environment. This memory is called eligibility trace ($z_t$) and is computed according to:

$$z_t = \lambda z_{t-1} + h_t(pre, post),$$

(6)

where $\lambda \in [0,1]$ denotes the decay rate (i.e., timescale) of the eligibility trace and $h_t(pre, post)$ the two-factor STDP update [8]. As long as the eligibility trace is nonzero, the synapse is eligible for a weight update. Hence, when a reward signal $\delta_t$ arrives, the weight gets updated according to:

$$\Delta w_t = \alpha \delta_t \cdot z_t,$$

(7)

where $\alpha$ denotes the learning rate[8].

**Correspondence between the device dynamics and the theory.** In Table 1 we summarize the list of correspondences between the physical properties of the device and the theoretical quantities of the 3-factor framework. In a simplified case the eligibility trace (Eq. 6) and the calculation of the weight update (Eq. 7), can be approximated in our hardware as follows: The intrinsically decaying conductance state ( $G_0$ ) models the local synaptic eligibility ( $z_t$) that is modified by an electrical signal ($V_{pulse}$), which represents $h_t(pre, post)$ in the simplified case where $h_t$ is equal to the pre-synaptic activity ($pre_t$). With a bias voltage ($V_{bias}$) the decay time constant ($\lambda$) of $G_0$ can be controlled (see Fig. 4c). The UV light intensity ($P_{opt}$) then corresponds to the reward signal, while the conductance-dependent photoresponse $\Delta G$ implements the processing of the reward, i.e., the local weight update ($\Delta w_t$) of each synapse according to Eq. 7. This calculation is visualized in Fig. 5d, which explains how the local weight update (labelled Theory) is calculated by our devices (labelled Device). The global reward implemented by the UV light as well as the local eligibility that is modelled by the decaying conductance state are indicated. The complete correspondence between the learning algorithm (Theory) and the device properties (Device) is summarized in Table 2.

| Theory | Device |
| --- | --- |
| $w_t$ | - |
| $h_t$ | $V_{pulse}$ |
| $\alpha\delta_t$ | $\sqrt{P_{opt}}$ |
| $z_t$ | $\sqrt{G_0}$ |
| $\lambda$ | $V_{bias}$ |
| $\Delta w_t$ | $\Delta G$ |

Table 2: Correspondence between three-factor learning in Eqs. (5)-(7) and our device properties.

It should, however, be noted that our devices are not yet fully capable of implementing the function of biological synapses in a three-factor learning model (see Table 3). In particular, to fully implement the R-STDP rule, the synapse model requires two distinct states: the non-volatile synaptic weight $w_t$ that accumulates the weight changes $\Delta w_t$, and a hidden eligibility $z_t$. With the proposed implementation, those two variables are intertwined as they are both a function of $G_0$, which is not a fully non-volatile state. Addressing this issue requires an in-depth understanding of the physical mechanisms governing the switching of our STO memristors, which will enable us to engineer next-generation components appropriately. Our measurements of the optical decay under different bias voltages (Fig. 4b) suggest that the volatility can be controlled in principle. Moreover, studies have shown that the introduction of oxygen blocking layers [53], [54] or electrodes [30] may lead to non-volatile (electrical) operation in practice. These experiments indicate that a non-volatile optical response could be possible. Besides this challenge, further work is needed to scale up our concept such as converting the devices to vertical structures. This is expected to reduce device to device variability thanks to a well-controlled electrode spacing and also allows for the creation of electro-optical crossbar arrays. Through an appropriate optical design plasmonic resonances of the top electrode can be harnessed to guide light to the active area at the electrode edges, which simultaneously increases the local electromagnetic field intensity ~~in the active area~~ and thereby limits optical crosstalk.

| Synaptic function | Implemented by our devices |
| --- | --- |
| (1) Dynamic memory state | ✓ |
| (2) State-dependent global modulation | ✓ |
| (3) Controllable decay time | ✓ |
| (4) Stable (non-volatile) weight | ✗ |

Table 3: Synaptic functions required for three-factor learning

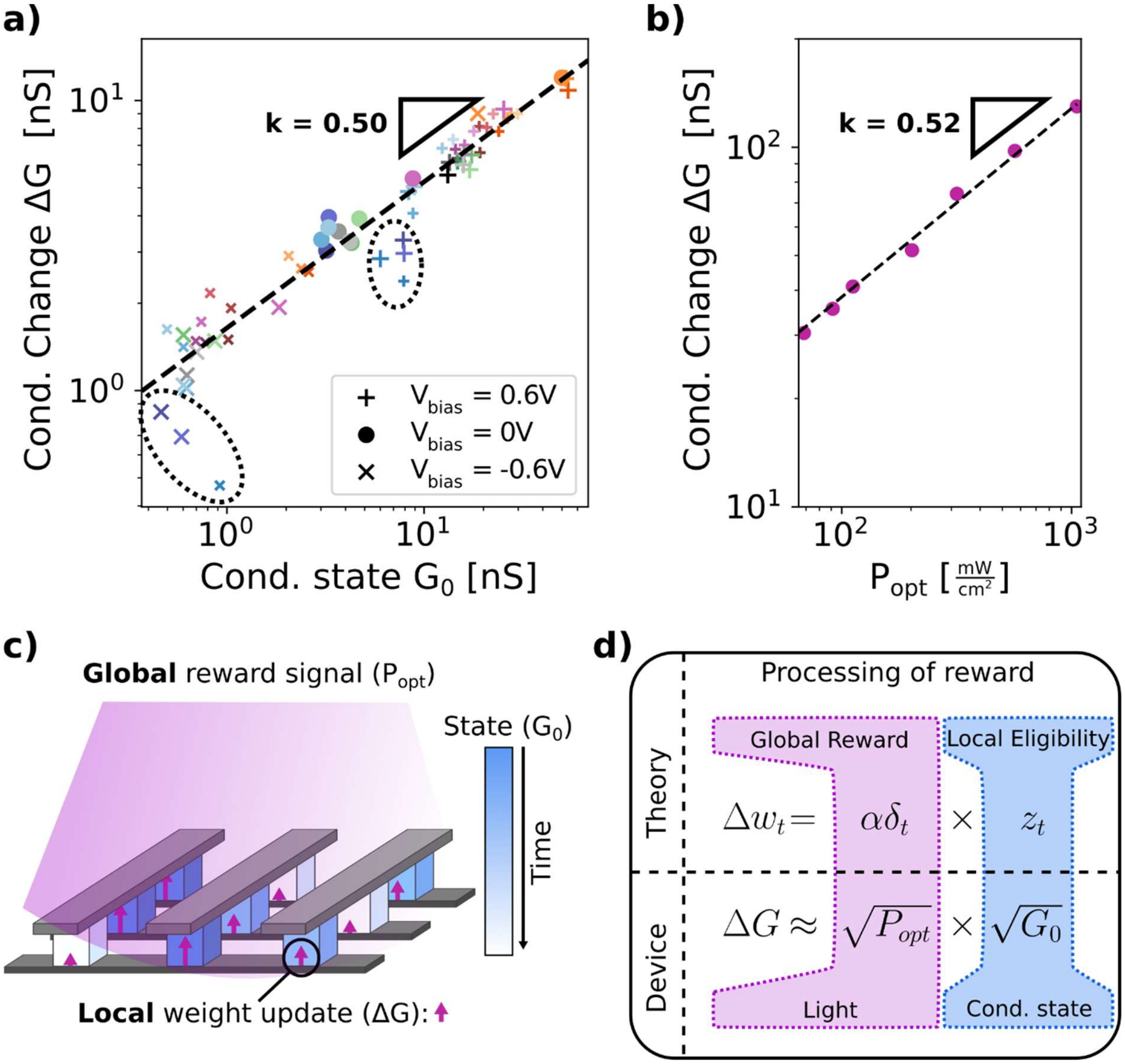

Figure 5: Summary of the photoresponse features and possible application of our STO devices in three-factor learning algorithms. (a) Optically induced conductance change $\Delta G^{(opt)}$ as a function of the conductance state $G_0$ for all measurements performed in this work. The dotted ovals mark outliers, whereas the dashed black line is a power law fit (exponent k=0.50) of the data without the outliers. Marker shape corresponds to $V_{bias}$, marker size to the employed measurement protocol and marker color to the timing relative to the SET pulse (see Supplementary Section S7, Fig. S4 for details). (b) $\Delta G^{(opt)}$ as a function of the optical power $P_{opt}$. The dashed black line is a power law fit (exponent k=0.52). (c) Artistic rendering of a possible hardware implementation of crossbar arrays with individual devices exhibiting different conductance states (eligibility, $G_0$). A global optical signal illuminated from top is envisioned to trigger local conductance updates that depend on $G_0$. (d) Computation of the theoretical weight update $\Delta w_t$ (upper part; labelled "Theory") and the mathematical formulation of the conductance-dependent photoresponse ($\Delta G$) extracted from our device (bottom part; labelled "Device").

## Conclusion and Outlook

In this work we demonstrated $SrTiO_3$-based electro-optical memristors that integrate several essential synaptic functions relevant for three-factor learning rules and other emerging neuromorphic applications: (1) a local transient memory state that (2) interacts with a global optical signal producing a state-dependent photoresponse and (3) whose temporal evolution is controllable.

Specifically, we showed that the device changes its internal conductance state upon optical or electrical excitation and then relaxes back toward its initial state with characteristic time constants ranging from milliseconds to hundreds of seconds (transient memory). These decay times can be tuned by applying a bias voltage, enabling the matching of the memory state's time-constant to specific tasks. Through a theoretical and experimental investigation, the physical origin of the decay was tracked to the back-migration of photogenerated oxygen vacancies. Crucially, the steadily decaying local conductance state influences the device's electrical response to a global optical signal: the larger the conductance state, the stronger the photoresponse, according to a well-defined square-root relation. In other words, because the conductance decays over time, optical signals that arrive earlier produce stronger photo-responses, effectively making the local electrical response sensitive to the timing of the global optical signal. This timing-dependent global-to-local conversion is central to three-factor learning rules, which are considered to be an important mechanism in biological learning. Finally, all these functions are achieved within a nanoscale footprint and at very low energy cost, meeting critical requirements for scalable neuromorphic hardware architectures.

Looking forward, two complementary research directions can be envisioned. On the device level, advancing the physical understanding of oxygen vacancy dynamics and trapping processes will be key to tailoring the balance between volatile and non-volatile operation. At a systems level, integrating STO memristors into crossbar arrays and testing them in hardware-implemented learning tasks will clarify their potential for bio-realistic computing. Array-level properties such as electrically and optically induced crosstalk will need to be investigated in detail to clarify the device's scale-up potential. If successful, these efforts could position electro-optical memristors as a compact and versatile platform enabling the implementation of complex learning functions in next-generation neuromorphic hardware.

## Contributions

C.W. developed the concept of the paper, fabricated and measured devices. He wrote the paper with input from all co-authors. H.H. provided the SEM cross section image and fabricated the SiN test structures. M.M. performed the DFT calculations and helped with the identification of the physical mechanism in the devices. T.Z. supported the development of the characterization setup and device measurements. K.P. assisted with fabrication. K.B. conducted the UV-Vis measurements of the SiN layer and helped with writing. G.B assisted with the bio-inspired learning algorithms and writing. M.L. supervised the project, gave inputs on the device working principle and helped with the writing and structuring of the paper. A.E. supervised the project and led the study with inputs on numerous topics including the fabrication/characterization of the devices and the writing of the paper. C.W. and A.E. conceived the project.

## Corresponding author

Correspondence to Christoph Weilenmann

## Acknowledgement

We would like to thank the Operations Team of the Binnig and Rohrer Nanotechnology Center, especially Antonis Olziersky, Roland Grundbacher, Ute Drechsler, Michael Stiefel, Steffen Reidt and Diana Davila for their continued help and sharing of their fabrication knowledge. Funding from the Werner Siemens Foundation (A.E., M.L., and M.M.), the SNSF Strategic Japanese-Swiss Science and Technology Program under project metacross (grant number 214068, C.W. and A.E.), the SNSF Sinergia project ALMOND (grant number 198612, M.L., and T.Z.), and the SNSF Advanced Grant project QuaTrEx (grant number 209358, M.L.) is acknowledged.